\documentclass{jpsj3}

\usepackage{txfonts}

\usepackage[dvipdfmx]{xcolor}

\newcommand{\TLnCO}{$Ln_2$CuO$_4$}
\newcommand{\TCud}{Cu3$d$}
\newcommand{\TCudxxyy}{Cu3$d_{x^2-y^2}$}
\newcommand{\TCudzzrr}{Cu3$d_{3z^2-r^2}$}
\newcommand{\TOp}{O2$p$}

\newcommand{\TDraft}{}

\newcommand{\TRevA}[1]{{#1}}
\newcommand{\TRevB}[1]{{#1}}
\newcommand{\TRevC}[1]{{#1}}
\newcommand{\TRevD}[1]{{#1}}
\newcommand{\TRevE}[1]{{#1}}
\newcommand{\TRevF}[1]{{#1}}
\newcommand{\TRevG}[1]{{#1}}
\newcommand{\TRevH}[1]{{#1}}

\title{Electronic State of T'-Pr$_{1.3-x}$La$_{0.7}$Ce$_x$CuO$_4$ ($x = 0.10$) Studied by Compton Scattering}

\author{
Takayuki Kawamata$^1$\thanks{E-mail: \TRevE{tkawamata@mail.dendai.ac.jp, Present address: Department of Natural Sciences, Tokyo Denki University, Tokyo 120-8551, Japan}}, 
Shin Saito$^1$, 
Naruki Tsuji$^2$, \\
\TRevB{Takuya Sumura$^3$,} 
Tadashi Adachi$^3$, 
Masatsune Kato$^1$, \\
Yoshiharu Sakurai$^2$, 
and 
Yoji Koike$^1$
}

\inst{
$^1$Department of Applied Physics, Tohoku University, Sendai 980-8579, Japan \\
$^2$Japan Synchrotron Radiation Research Institute (JASRI), SPring-8, Sayo, Hyogo 679-5198, Japan \\
$^3$Department of Engineering and Applied Sciences, Sophia University, Tokyo 102-8554, Japan
}

\abst{
We have performed Compton scattering \TRevF{measurements on} as-grown and reduced single crystals of the electron-doped T'-cuprate Pr$_{1.3-x}$La$_{0.7}$Ce$_x$CuO$_4$ ($x = 0.10$) to investigate 
\TRevF{the effect of reduction annealing on} 
the electronic state. 
\TRevF{The obtained results} have revealed that 
the numbers of electrons in the {\TOp} orbital in the Zhang--Rice singlet band, 
the {\TCudxxyy} orbital, and the {\TCudzzrr} orbital are increased by reduction annealing.  
The increase in the number 
\TRevF{of} {\TCudzzrr} \TRevF{electrons} 
suggests the existence of 
{\TCudzzrr} \TRevF{holes in} 
the as-grown single crystal. 
This is \TRevD{spectroscopic evidence} of local hole doping into the {\TCudzzrr} orbital around the excess oxygen, 
as suggested \TRevF{by} transport \TRevF{measurements} [T. Adachi \textit{et al.}, J. Phys. Soc. Jpn. \textbf{82}, 063713 (2013)]. 
}

\begin{document}
\maketitle

\TDraft

\section{Introduction}
\TRevF{Among the cuprates,} 
{\TLnCO} ($Ln$: lanthanide elements) with the Nd$_2$CuO$_4$-type (so-called T'-type) structure
\TRevF{is particularly interesting, since} 
superconductivity appears without carrier doping \TRevD{in adequately reduced} thin films 
\cite{Tsukada:SSC133:2005:427,Matsumoto:PC469:2009:924,Matsumoto:PRB79:2009:100508} 
and polycrystalline bulk samples \cite{Asai:PC471:2011:682,Takamatsu:APE5:2012:073101} of undoped mother \TRevF{compounds.} 
Moreover, in single crystals of electron-doped (Ce-doped) T'-(Pr,La)$_{2-x}$Ce$_x$CuO$_4$\cite{Horio:NC7:2016:10567,Adachi:CM2:2017:23} 
\TRevD{and T'-Pr$_{2-x}$Ce$_x$CuO$_4$,\cite{Brinkmann:PRL74:1995:4927}} 
superconductivity has been observed extensively in the underdoped region down to \TRevD{$x = 0.04$--$0.05$.}
\TRevF{These results suggest} 
that undoped T'-{\TLnCO} is not a Mott insulator but a strongly correlated metal.  
To investigate the electronic state of undoped (Ce-free) \TRevH{superconductors}, 
various experiments have been carried out, 
\cite{Adachi:JPSJ82:2013:063713,Ohashi:JPSJ85:2016:093703,Adachi:JPSJ85:2016:114716,Horio:NC7:2016:10567,Fukazawa:PC541:2017:30,Ohnishi:JPSJ87:2018:043705,Asano:JPSJ87:2018:094710,Kawamata:JPSJ87:2018:094717,Sunohara:JPSJ89:2020:014701,Lee:JPSJ89:2020:073709} 
\TRevD{and} two models of the electronic state have been proposed.  
One is a strongly correlated metallic state without a charge-transfer (CT) gap, 
\cite{Adachi:JPSJ82:2013:063713} 
which is the energy gap between the upper Hubbard band (UHB) of the {\TCudxxyy} orbital and the {\TOp} band, as shown in Fig. 1(a). 
According to this model, electron and hole carriers appear in undoped T'-{\TLnCO} owing to the overlap of the UHB of the {\TCudxxyy} orbital 
and \TRevE{the} Zhang--Rice singlet band 
\TRevE{composed of holes in the {\TOp} orbital and electrons in the \TRevH{lower Hubbard band (LHB)} of 
the {\TCudxxyy} orbital}. 
On the basis of this model, the preceding spectroscopic results 
\cite{Moritz:NJP11:2009:093020,Schmitt:PRB83:2011:195123,Basak:PRB85:2012:075104} 
of the finite CT gap in the reduced T'-cuprates are explained \TRevE{as being due to} the insufficient reduction of excess oxygen at the apical site. 
The other is a strongly correlated metallic state with a finite CT gap, 
as shown in \TRevD{Fig. 1(d)}.\cite{Horio:NC7:2016:10567,Ohnishi:JPSJ87:2018:043705} 
According to this model, electron carriers are doped into the UHB of the {\TCudxxyy} orbital owing to oxygen defects induced by reduction annealing. 

\begin{figure}[b]
	\begin{center}
		\includegraphics[scale=0.13]{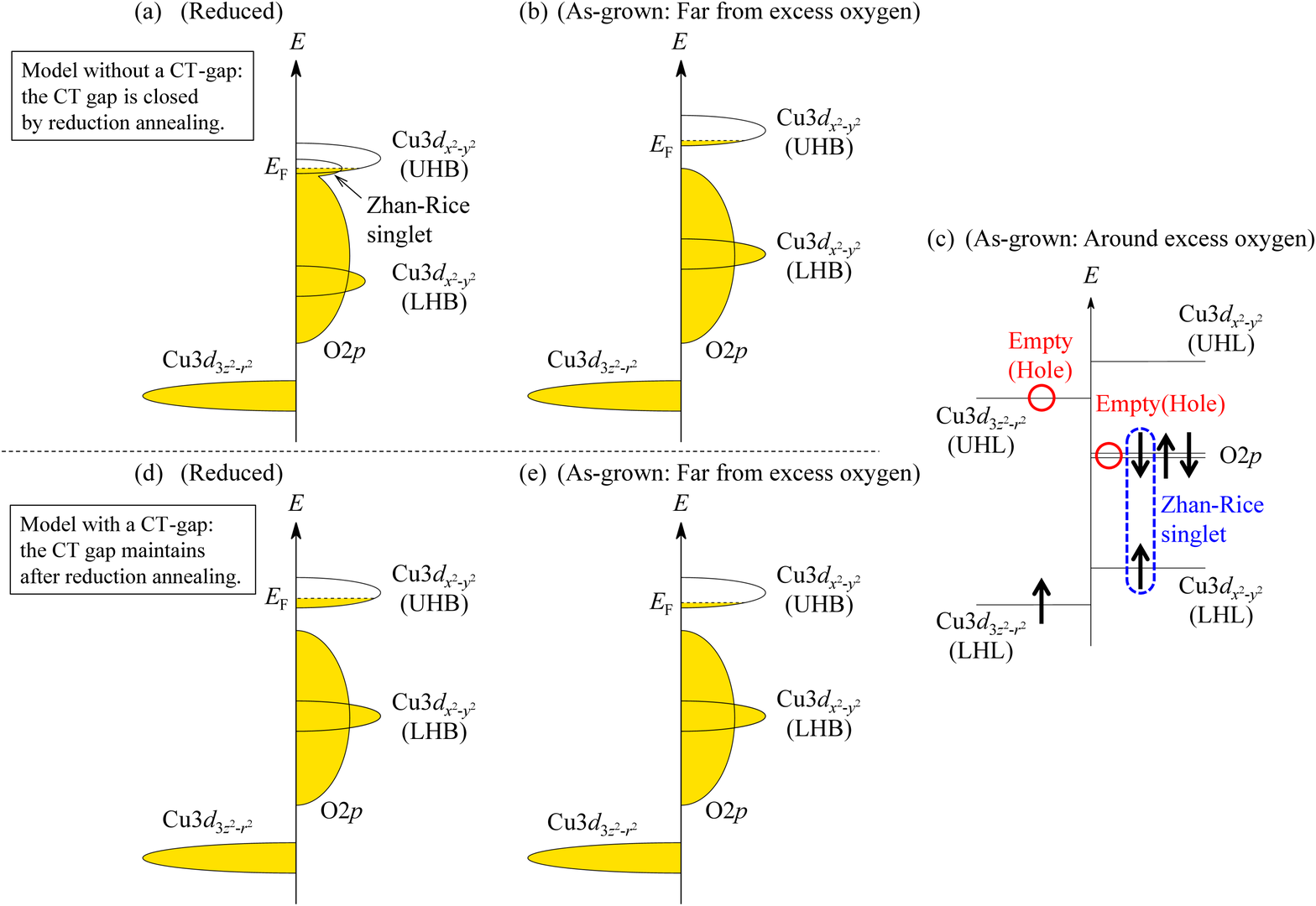}
		\caption{
			(Color online) \TRevE{Schematic} electronic structures of 
			T'-Pr$_{1.3-x}$La$_{0.7}$Ce$_x$CuO$_4$ with $x = 0.10$ based on the 
			strongly correlated metallic states 
			(a)\TRevD{(b)} without a charge-transfer (CT) gap
			\cite {Adachi:JPSJ82:2013:063713,Adachi:JPSJ85:2016:114716,Adachi:CM2:2017:23} 
			and (d)(e) with a finite CT gap. 
			\cite{Horio:NC7:2016:10567,Ohnishi:JPSJ87:2018:043705} 
			\TRevE{(a)(d) Electronic structures in the reduced single crystal. 
			(b)(e) Electronic structures far from excess oxygen in the as-grown single crystal.  
			(c) Local electronic structure around excess oxygen in the as-grown single crystal.
			\cite{Adachi:JPSJ82:2013:063713,Adachi:CM2:2017:23}}
			Two holes (red circles) are \TRevD{generated} 
			by the ionization of one excess oxygen.  
			UHB and LHB mean the upper and lower Hubbard bands, respectively. 
			\TRevD{In the case of \TRevE{the} local electronic structure of (c), 
			UHL and LHL mean the upper and lower Hubbard levels, respectively.} 
			\TRevD{Here, we consider a rigid-band-like picture 
			in which the band structure is not changed by carrier doping.} 
		}
		\label{Fig:Band}
	\end{center}
\end{figure}

Compton scattering \TRevF{measures} 
the electron momentum density (EMD) of a sample, 
which reflects the symmetry of \TRevF{electron} orbitals.  
In hole-doped La$_{2-x}$Sr$_x$CuO$_4$ with the K$_2$NiF$_4$ \TRevD{structure, the Compton} scattering experiment has revealed that the \TRevB{EMDs} along the [100] and [110] directions are changed 
by hole doping. \cite{Sakurai:S332:2011:698} 
\TRevA{By calculations of the molecular orbital and the band structure, this result has been explained as being due to the hole doping into the {\TOp} orbital in the Zhang--Rice singlet band in the underdoped region and that into the {\TCudxxyy} and {\TCudzzrr} orbitals in the overdoped region.}  

In T'-{\TLnCO}, 
the \TRevF{EMDs} obtained from the Compton scattering experiment \TRevF{are} expected to 
reveal orbitals\TRevF{,} in which the number of electrons \TRevF{is changed} 
by reduction annealing.  
{
\TRevC{Since \TRevF{the} undoped (Ce-free) superconducting samples of T'-{\TLnCO} are polycrystalline \TRevF{or} filmy, 
it is difficult to obtain orbital information from the Compton scattering experiment.} 
Therefore, we \TRevE{used} single crystals of lightly electron-doped T'-Pr$_{1.3-x}$La$_{0.7}$Ce$_x$CuO$_4$ (PLCCO) with $x = 0.10$, where superconductivity appears upon strong reduction annealing, 
namely, the so-called protect annealing. \cite{Adachi:JPSJ82:2013:063713}.  
Through protect annealing, 
the \TRevE{single crystal} of $x=0.10$ becomes metallic and exhibits superconductivity, 
whereas the conventional annealing in flowing Ar gas is insufficient to \TRevD{induce} superconductivity. \cite{Sun:PRL92:2004:047001}
\TRevD{In the insufficiently reduced} \TRevE{single crystal}, 
the temperature dependence of the $ab$-plane electrical resistivity exhibits a log$T$-dependent upturn at low temperatures.  
Moreover, negative magnetoresistance is observed at the corresponding temperature range.
These results suggest the occurrence of Kondo scattering, 
which is presumably due to a small amount of residual excess oxygen.\TRevD{\cite{Sekitani:JPCS63:2002:1089,Sekitani:PRB67:2003:174503}}
As shown in Fig. 1(c), 
the excess oxygen in the as-grown T'-cuprates \TRevE{is inferred to} \TRevD{generate} two holes in the CuO$_2$ plane,
 and one of them resides in the \TRevD{upper Hubbard level (UHL)} of the {\TCudzzrr} orbital, giving rise to \TRevD{an unpaired} spin in the \TRevH{lower Hubbard level (LHL)} of the {\TCudzzrr} orbital. 
This \TRevE{unpaired spin is regarded as} a scatterer of electrons that brings about the Kondo effect, although no spectroscopic evidence has yet been obtained. 
The Compton scattering experiment using single crystals of T'-PLCCO with $x=0.10$ is expected to provide deep insight into the change in the \TRevF{electron} orbitals caused by reduction annealing, which would be of great help in understanding the electronic state of the undoped superconductivity. 
}
Accordingly, we have performed Compton scattering measurements 
on as-grown and reduced single crystals of T'-PLCCO with $x = 0.10$. 

\section{Experimental}
Single crystals of T'-PLCCO with $x = 0.10$ were grown by the traveling-solvent floating-zone method. \cite{Adachi:JPSJ82:2013:063713} 
The reduced single crystal was obtained 
by protect annealing,\cite{Adachi:JPSJ82:2013:063713} 
and it showed superconductivity at temperatures below $\sim$27 K. 
The Compton scattering experiment was carried out at room temperature for the as-grown and reduced single crystals of T'-PLCCO with $x=0.10$ at the BL08W beamline of SPring-8, Japan. 
\TRevF{Ten} Compton profiles were measured at even intervals between the [100] and [110] directions to obtain two-dimensional (2D) EMDs that \TRevF{represent} one-dimensional integrals along the $c$-axis of three-dimensional EMDs.  

\section{Results and Discussion}

\begin{figure}[b]
	\begin{center}
		\includegraphics[scale=0.75]{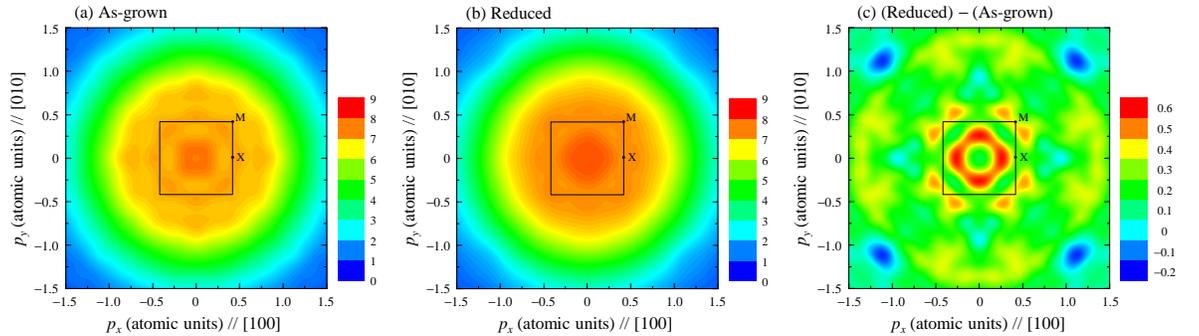}
		\caption{
			(Color online) \TRevD{Two-dimensional electron momentum densities (2D EMDs) of 
			the (a) as-grown and 
			(b) reduced single crystals of T'-Pr$_{1.3-x}$La$_{0.7}$Ce$_x$CuO$_4$ with $x = 0.10$.
			(c) \TRevE{Difference} 2D EMD of the reduced single crystal 
			obtained after subtracting the 2D EMD of the as-grown single crystal.}  
			$p_x$ and $p_y$ are electron momentums along the $a$- and $b$-axes, respectively.
			The black square indicates the first Brillouin zone. 
			}
		\label{Fig:DiffEMDs}
	\end{center}
\end{figure}

Figures \ref{Fig:DiffEMDs}(a) and 2(b) show 2D EMDs of the as-grown and reduced 
single crystal\TRevE{s} of T'-PLCCO with $x = 0.10$, respectively.  
To clearly see the effect of reduction annealing, 
the difference 2D EMD of the reduced single crystal 
obtained after subtracting the 2D EMD of the as-grown single crystal 
\TRevE{is shown in Fig. \ref{Fig:DiffEMDs}(c)}. 
It is found that the \TRevG{difference} 2D EMD shows the maximum around the X point.  
\TRevA{A similar behavior has been observed in the underdoped and optimally doped regions of La$_{2-x}$Sr$_x$CuO$_4$.} 
\cite{Sakurai:S332:2011:698} 
The increase in the 2D EMD around the X point 
has been explained as being due to the increase in the number of electrons in the Zhang--Rice singlet band, using the molecular orbital calculation. 
\cite{Sakurai:S332:2011:698} 
\TRevD{Since the calculation on La$_{2-x}$Sr$_x$CuO$_4$ is based on {\TOp} and {\TCudxxyy} orbitals in the CuO$_2$ plane, the results can be applied to T'-{\TLnCO}.}
Considering that the {\TOp} and {\TCud} orbitals tend to 
\TRevB{mainly} 
contribute to the 2D EMD in the first Brillouin zone (BZ) and high BZs, respectively, 
\cite{Sakurai:S332:2011:698}  
this result indicates that the number of electrons is increased in the {\TOp} orbital in the Zhang--Rice singlet band \TRevD{by reduction annealing}. 
Furthermore, it is \TRevF{also} found that the \TRevG{difference} 2D EMD increases around the M point. 
A similar behavior has been observed in the overdoped region of La$_{2-x}$Sr$_x$CuO$_4$ and has been explained as being due to the increase in the number of electrons in the {\TOp} and {\TCudzzrr} orbitals. 
\cite{Sakurai:S332:2011:698} 
\TRevF{In addition}, it is found that the \TRevG{difference} 2D EMD at around $(p_x, p_y) = (0, \pm1.3)$ and $(\pm1.3, 0)$ slightly increases, meaning that the number of electrons in the UHB of the {\TCudxxyy} orbital increases. 
Therefore, these results indicate that the numbers of electrons in \TRevE{the} 
{\TOp}, {\TCudxxyy}, and {\TCudzzrr} orbitals are increased by reduction annealing. 
\TRevD{The removal of excess oxygen results in electron doping. 
Therefore, 
it is reasonable that the number of electrons in these orbitals near the Fermi level increases.}

The observation of electron doping into the {\TCudzzrr} orbital by reduction annealing is a notable feature in these results. 
For both electronic structure models without and with \TRevE{a} CT gap shown 
in Figs. 1(a) and \TRevF{1(d)}, 
respectively, the {\TCudzzrr} band is completely filled with electrons 
\TRevE{in the reduced single crystal}. 
\TRevD{In the as-grown single crystal, therefore, 
the local electronic structure around excess oxygen must be the one shown in Fig. 1(c), in which a hole resides on the UHL of the {\TCudzzrr} orbital, for both models.
This is also a new suggestion for the electronic structure model with \TRevE{a} CT gap in the reduced \TRevE{single crystal}}. 
\TRevE{By the insufficient reduction of excess oxygen, on the other hand, 
the UHL of the {\TCudzzrr} orbital is not fully filled, 
so that unpaired spins in the LHL of the {\TCudzzrr} orbital bring about Kondo scattering, 
as observed 
in the insufficiently reduced T'-PLCCO with $x=0.10$ \cite{Adachi:JPSJ82:2013:063713}.}
Accordingly, this result is \TRevD{spectroscopic evidence} of the orbital state around the excess oxygen shown in Fig. 1(c).
In the hole-doped cuprates, \TRevF{the} effects of the {\TCudzzrr} orbital on the electronic state have been discussed \cite{Sakakibara:PRL105:2010:057003,Sakurai:S332:2011:698}.
\TRevF{To the best of our knowledge}, 
it is the first observation of \TRevF{the} effects of the {\TCudzzrr} orbital 
on the electronic state of the electron-doped T'-cuprates. 
In the mother compound exhibiting undoped superconductivity, it is possible that a similar change in the electronic state \TRevF{takes place} after reduction annealing. 

\TRevD{Here, the increase in the numbers of electrons in the {\TOp} and {\TCudxxyy} orbitals is discussed on the basis of the rigid-band picture. 
In as-grown T'-PLCCO with $x=0.10$, as shown in Fig. 1(c), 
there is a hole in the {\TOp} orbital around excess oxygen.} 
The removal of excess oxygen results in the increase in the number of electrons in the {\TOp} orbital, 
which is consistent with these results. 
How is the increase in the number of electrons in the {\TCudxxyy} orbital explained? 
\TRevD{First, we discuss the case of the model without a CT gap. 
In as-grown \TRevE{T'-PLCCO}, 
the electronic state around excess oxygen is represented by Fig. 1(c) and that far from excess oxygen is represented by Fig. 1(b). 
In the case of Fig. 1(b), the so-called antiferromagnetic pseudogap opens on the Fermi surface, 
as observed by angle-resolved photoemission spectroscopy (ARPES) experiments,\cite{Horio:NC7:2016:10567} 
owing to \TRevE{the} localization of carriers around the excess oxygen 
and therefore as-grown \TRevE{T'-PLCCO} becomes insulating. 
The removal of excess oxygen leads to the delocalization of carriers, 
\TRevE{the} destruction of the antiferromagnetic order, and the gap closing on the Fermi surface. 
These change the electronic structure to the one shown in Fig. 1(a) throughout the sample. 
That is, the UHB of the {\TCudxxyy} orbital and the Zhang--Rice singlet band overlap, 
increasing the number of electrons in the UHB of the {\TCudxxyy} orbital.} 
The results of ARPES experiments in reduced T'-Nd$_{2-x}$Ce$_x$CuO$_4$ have suggested the existence of a quasiparticle band in the CT gap. \cite{Armitage:PRL88:2002:257001,Weber:NP6:2010:574} 
\TRevF{The quasiparticle band probably corresponds to the Zhang--Rice singlet band \TRevD{in Fig. 1(a).}} 

\TRevD{Next, we discuss the case of the model with a CT gap.
In as-grown \TRevE{T'-PLCCO}, 
the electronic states around and far from excess oxygen are represented by Figs. 1(c) and 1(e), respectively, as mentioned above.
Through reduction annealing, electrons are doped into the UHB of the {\TCudxxyy} orbital owing to oxygen \TRevE{defects} in the blocking layer and/or the CuO$_2$ plane, leading to the electronic structure shown in Fig. 1(d). 
The quasiparticle band in the CT gap observed in the reduced T'-Nd$_{2-x}$Ce$_x$CuO$_4$ by ARPES 
\cite{Armitage:PRL88:2002:257001,Weber:NP6:2010:574} cannot be explained by the present model on the basis of the rigid-band picture. 
Although the band structure would be changed in a rather complicated form with carrier doping, 
the quasiparticle band would be \TRevF{a} mixed band between the {\TCudxxyy} and {\TOp} orbitals in the CuO$_2$ plane because of a strong covalent state. 
Therefore, the electrons doped by reduction annealing would reside in the mixed band.} 
\TRevG{Accordingly, the these results of electron doping into the UHB of the {\TCudxxyy} orbital 
\TRevF{through} reduction annealing are understandable 
\TRevD{in both models without and with \TRevE{a} CT gap.}}

\section{Summary}
We have performed Compton scattering \TRevF{measurements} on as-grown and reduced single crystals of T'-PLCCO with $x = 0.10$ to obtain information on the electronic state. 
From \TRevF{10} Compton profiles between the [100] and [110] directions, we have obtained 2D EMDs of the as-grown and reduced single crystals. 
The \TRevG{difference} 2D EMD between \TRevE{the} as-grown and reduced single crystals has revealed that the 2D EMD around the X point, $(p_x, p_y) = (0, \pm1.3)$ and $(\pm1.3, 0)$, and the M point becomes large by the reduction annealing, indicating that the numbers of electrons in the {\TOp} orbital in the Zhang--Rice singlet band, the {\TCudxxyy} orbital, and the {\TCudzzrr} orbital increase. 

{The increase in the number of electrons in the {\TCudzzrr} orbital after reduction annealing suggests hole doping into the {\TCudzzrr} orbital around the excess oxygen in the as-grown single crystal, which is \TRevD{spectroscopic evidence} of our proposed model shown in Fig. 1(c).
On the other hand, the increase in the number of electrons in the Zhang--Rice singlet band
is understandable in both models without and with \TRevE{a} CT gap. 
These results would contribute to the understanding of the electronic state of the undoped superconductivity in the mother compound of T'-cuprates.
}

\begin{acknowledgment}
This work was supported by JSPS KAKENHI Grant Numbers \TRevB{23108004, 17H02915, and 19H01841}. 
\TRevB{The Compton scattering experiment was performed with the approval of JASRI (Proposal No. 2017B1222).}
\end{acknowledgment}



\end{document}